# QLink: Quantum-Safe Bridge Architecture for Blockchain Interoperability

Joao Vitor Barros Da Silva, Arsh Gupta, Madhusudan Singh* and Irish Singh

Penn State University, University Park, PA, USA

Jmb9000@psu.edu, abg6210@psu.edu, *msingh@psu.edu iks5301@psu.edu

*Corresponding Author

***Abstract:*** Secure interoperability across heterogeneous blockchains remains one of the most pressing challenges in Web3, with existing bridge protocols vulnerable to both classical exploits and emerging quantum threats. This paper introduces QLink, a quantum-safe Layer-3 interoperability protocol that integrates post-quantum cryptography (PQC), quantum key distribution (QKD), and hardware security modules (HSMs) into a unified validator architecture. To our knowledge, QLink is the first interoperability framework to combine these mechanisms to secure validator communication, proof aggregation, and key management. Validators exchange encryption keys through QKD channels, achieving information-theoretic security against interception, while cross-chain proofs are generated and aggregated with NIST-standardized PQC algorithms. Private keys remain sealed inside HSM enclaves, mitigating the risk of theft or leakage. Deployed as a dedicated Layer-3 protocol, QLink operates independently of Layer-1 and Layer-2 chains, providing a scalable, decentralized foundation for secure cross-chain messaging and asset transfer. Experimental evaluation using network simulations demonstrates that validator communication overhead remains sub-second, while security guarantees extend beyond current bridge architectures to resist both classical and quantum adversaries. By addressing today's vulnerabilities and anticipating future quantum threats, QLink establishes a practical and future-proof pathway for blockchain interoperability.

## 1. Introduction

Decentralized finance (DeFi) is a new set of financial apps that use blockchain technology and don't need banks or brokerages to work. It lets people lend, borrow, trade, and earn interest through smart contracts that run on decentralized networks, mostly Ethereum and other programmable blockchains. DeFi replaces traditional banks with open, permissionless protocols. This makes it easier for more people to access financial services, reduces the need for trust-based systems, and makes financial services available to everyone around the world. However, it also creates new problems, such as security holes, regulatory uncertainty, and market volatility.in fig. 1 has represented an overview of DeFi System.

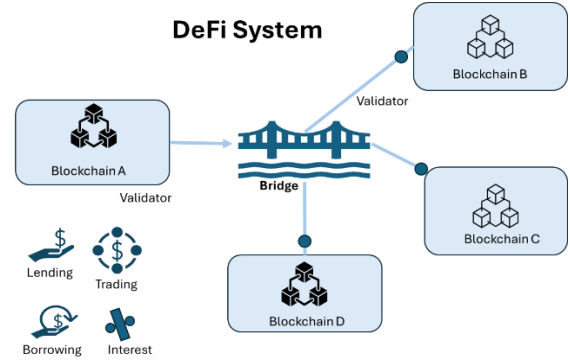

Fig.1. Decentralized Finance (De-Fi) System

In De-Fi System, Blockchain bridges are currently among the most vulnerable yet indispensable components of Web3 infrastructure. Designed to enable interoperability between heterogeneous blockchains, such as transferring assets or data between Bitcoin, Ethereum, and emerging Layer-2 networks, bridges allow users to move value across ecosystems that otherwise operate in isolation.

At a high level, bridges do not literally move tokens from one blockchain to another. Instead, they lock or escrow assets on the source chain and mint a *synthetic* or *wrapped* representation of those assets on the destination chain. When users withdraw, the wrapped tokens are burned, and the corresponding native assets are released back to the original chain. This mechanism allows users to transfer value between chains without relying on centralized exchanges, reducing friction and enhancing decentralization.

However, the design of bridges varies significantly along a spectrum of centralization. Trusted (centralized) bridges depend on a single custodian or a small federation of validators who control the locked funds, such as Wrapped Bitcoin (custodian: BitGo) or Wormhole's 19 "guardians." While efficient, these models introduce counterparty risk, since compromising a few validator keys can grant attackers full control. Trustless (decentralized) bridges, in contrast, rely on cryptographic proofs and smart contracts. Users deposit assets into an on-chain contract, and relayers propagate cryptographic proofs of these deposits to contracts on another chain. Once verified, new wrapped assets are minted automatically. This removes centralized custody but introduces new attack surfaces in smart contract logic and validator communication.

Despite their utility, cross-chain bridges have proven to be the single largest attack vector in decentralized finance (DeFi). According to DeFiLlama, bridge exploits have accounted for over $2.8 billion in losses, representing roughly 40% of all funds stolen in Web3 to date. Major incidents such as the Ronin Bridge ($625M, 2022), Wormhole ($320M, 2022), and Multichain ($130M, 2023) highlight how validator key theft, smart contract bugs, and insufficient monitoring can lead to catastrophic breaches.

Three recurring weaknesses drive most of these exploits:

1. Validator key theft or mismanagement, where compromised private keys enable attackers to drain entire bridges.
2. Forged or malicious proofs, where flaws in verification logic allow unauthorized minting or withdrawal of assets; and
3. Lack of forward security, as classical cryptographic algorithms (e.g., ECDSA, Ed25519) will eventually be broken by quantum computers, enabling "harvest-now, decrypt-later" attacks on recorded bridge communications.

The blockchain industry therefore lacks a quantum-resilient interoperability layer, one that can secure validator communication, proof verification, and key custody both against today's classical exploits and tomorrow's quantum adversaries.

To address these challenges, we introduce QLink, a novel Layer-3 quantum-safe interoperability protocol. Unlike approaches that patch existing bridge designs, QLink reimagines cross-chain architecture from the ground up by integrating two technologies. Quantum Key Distribution (QKD) provides a communication layer with information-theoretic security, enabling tamper-evident key exchange, mitigating man-in-the-middle attacks, and ensuring perfect forward secrecy. Post-Quantum Cryptography (PQC), using NIST-standardized algorithms like Dilithium and Falcon, enforced within Hardware Security Modules (HSMs), ensures validator keys never leave secure hardware and cross-chain proofs cannot be forged, even by quantum adversaries. Deployed as a dedicated Layer 3 protocol, QLink functions as a universal routing and validation layer above existing blockchains, providing modular connectivity, security isolation, and scalability. This makes QLink not only resistant to today's bridge exploits but also future proof against quantum threats.

## 2. Related Work

Research on blockchain interoperability and quantum-safe security has progressed along several directions. We group prior work into two categories: classical bridge designs and quantum-secure blockchain proposals. Classical Bridges. Current cross-chain protocols (e.g., Polygon PoS Bridge, Chainlink CCIP, Polkadot XCMP) rely on validator committees, threshold multisignatures, or multi-party computation (MPC) to coordinate lock-and-mint transfers. While widely deployed, these designs are vulnerable to validator key compromise and proof forgery, leading to multibillion-dollar bridge exploits such as Ronin (2022) and Wormhole (2022). These works demonstrate the limitations of classical cryptography and the need for stronger validator isolation.

Quantum-Secure Blockchain Proposals. Several studies have proposed integrating quantum-safe primitives into blockchain itself. For instance, Li et al. [10] and Alharbi et al. [9] explored the use of QKD and PQC for secure blockchain consensus. Mousa et al. [11] introduced QuantumShield-BC, a quantum-secure blockchain that natively integrates QKD and PQC at the ledger layer. Similarly, Thapliyal & Srinivasan [2] studied synergies between blockchain and quantum computing. While these works demonstrate feasibility of quantum-safe primitives in distributed systems, they focus on designing new blockchains, not on securing cross-chain bridges.

QLink's Differentiation. Unlike prior approaches, QLink is the first interoperability framework that:

- Targets bridges rather than standalone blockchains.
- Operates as a dedicated Layer-3 protocol, decoupled from Layer-1 consensus.
- Integrates QKD, PQC, and HSMs in validator communication, proof aggregation, and key custody simultaneously.

*Table 1. Summary of related work and differentiation of QLink*

| System / Protocol | Focused Key Mechanisms | Quantum-Safe? | Limitations |
|---|---|---|---|
| Polygon PoS Bridge [10] | Ethereum ↔ Polygon assets<br>Validator multisigs, Asset bridge | No | Susceptible to validator key theft; not quantum-resilient |
| Chainlink CCIP [11] | General interoperability<br>Oracle networks + MPC<br>Messaging + assets | No | Classical crypto only; MPC leakage risks |
| Polkadot XCMP [12] | Native parachain comms<br>Shared relay chain<br>Within Polkadot | No | Restricted to Polkadot ecosystem |
| Li et al. (2021) [13] | Secure blockchain comms | Partial | Focuses on intra-chain |
| | QKD-based channels Blockchain comms | | comms, not cross-chain bridges |
| Mousa et al., 2023 [14] | L1 blockchain design QKD + PQC Entire blockchain | Yes | Requires new blockchain; not an interoperability solution |
| Qlink [15] [16] | Cross-chain interoperability QKD + PQC + HSM Layer-3 bridge | Yes | Requires validator QKD hardware, but deployable over existing chains |

### A. Security Threat Model

We consider both classical blockchain adversaries and quantum-capable adversaries. By classifying them into two primary groups Classical Blockchain Adversaries and Quantum-Capable Adversaries, Fig. 6 shows the changing security risks that blockchain systems must contend with. Current threats like validator compromise and key theft, proof forgery and replay attacks, and communication interception where data is intercepted now and decrypted later are represented by classical adversaries. These dangers take advantage of flaws in conventional cryptography and system vulnerabilities. However, future attackers with quantum computing power pose a growing threat in the form of Quantum-Capable adversaries. They could launch availability and misbehavior attacks against validators or perform quantum-capable signature forgeries, compromising network dependability and consensus. In order to guarantee long-term security and trust, the model emphasizes the critical need to fortify blockchain systems with quantum-safe cryptography and improved validator protections. Threats are grouped into five main categories as shows in Fig. 2:

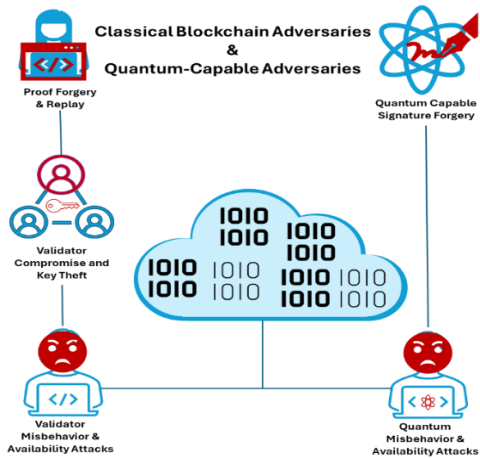

Fig. 2. Security Threat Models in Blockchain Interoperability

### B. Validator Compromise and Key Theft

Validator compromise and key theft happen when an attacker gets hold of a validator's private key. This lets them create fake signatures and steal bridge contracts, as happened in Ronin 2022. Malware, insider threats, or direct cryptographic key theft can all allow enemies to get into validator machines. QLink reduces this risk by making sure that all validator private keys stay sealed in HSM enclaves and are never exposed to software. It also requires hardware-enforced authorization for signatures, which stops direct theft and makes the attack surface much smaller than it would be with hot-wallet or MPC-based bridge architectures.

Validator compromises and key theft have following key points:

- **Threat:** In classical bridges, compromise of a validator's private key allows attackers to forge signatures and drain bridge contracts (e.g., Ronin 2022).
- **Adversary Capability:** Access to validator machines through malware, insider attacks, or theft of cryptographic keys.
- **QLink Defense:** All validator private keys remain sealed within HSM enclaves, never exposed to software. Signatures require hardware-enforced authorization. This eliminates direct theft and reduces attack surface compared to hot-wallet or MPC-based bridges.

### C. Proof Forgery and Replay

Malicious validators use fake proofs or replay old ones to start unauthorized mint or release events, as seen in attacks like Wormhole 2022. Adversaries can do this by controlling a small group of validators or taking advantage of weak proof-verification logic. QLink protects itself from these threats by using post-quantum cryptographic digital signatures (like Dilithium and Falcon) to combine proofs. These signatures are what bridge contracts check. Before reaching consensus, validators check SPV or light-client proofs on their own. The Byzantine Fault Tolerant consensus model needs at least ≥2f+1 honest validators to sign, which stops a small group of validators from working together.

Proof forgery and reply have 3 challenges shown below:

- **Threat:** Malicious validators may attempt to inject fake proofs or replay old proofs to

trigger unauthorized mint/release events (e.g., Wormhole 2022).

- *Adversary Capability:* Control of a minority subset of validators or exploiting weak proof-verification logic.
- *QLink Defense:* QLink employs PQC digital signatures (e.g., Dilithium, Falcon) for proof aggregation, verified by bridge contracts. Validators independently verify SPV or light-client proofs before consensus. Byzantine Fault Tolerant consensus requires ≥2f+1 honest validators to sign, preventing minority collusion.

### D. Communication Interception and Harvest-Now, Decrypt-Later

Attacks that intercept and harvest communications now and decrypt them later are a long-term threat. In these attacks, enemies record validator communications today with the plan to decrypt them later when quantum computers can break RSA or ECDSA. These kinds of enemies are like passive eavesdroppers who have a lot of storage space and might be able to decrypt quantum data in the future. To counter this, QLink protects all validator communications with Quantum Key Distribution (QKD)-generated one-time pad (OTP) encryption and VMAC authentication. This ensures information-theoretic confidentiality and makes recorded ciphertexts permanently useless, even to attackers who can use quantum computers in the future.

The communication interpretation and harvest - now, decrypt later show the challenges as follows:

- *Threat:* An adversary records validator communication today with the goal of decrypting it later once quantum computers can break RSA/ECDSA.
- *Adversary Capability:* Long-term passive eavesdropper with storage and future quantum decryption power.
- *QLink Defense:* All validator communication uses QKD-generated one-time pad (OTP) encryption with VMAC (Vectorized Message Authentication Code) authentication. This achieves information-theoretic confidentiality: recorded ciphertexts are useless even against future quantum adversaries.

### E. Validator Misbehavior and Availability Attacks

Validator misbehavior and availability attacks occur when validators go offline, fail to deliver QKD keys, or attempt double-signing, and adversaries may further disrupt the system through denial-of-service attacks on the QKD plane. Such threats arise from malicious insiders or external attackers targeting validator availability. QLink addresses these risks through protocol-level penalties such as slashing or disqualification for double-signing or key delivery failures, while its hub-and-spoke QKD architecture provides redundant key paths to sustain operations and mitigate targeted outages.

The validator misbehavior and availability attack shows the issues as follows:

- **Threat:** Validators may go offline, fail to deliver QKD keys, or attempt double-signing. Adversaries may also launch denial-of-service attacks on the QKD plane.
- **Adversary Capability:** Malicious insiders or external attackers targeting validator availability.
- **QLink Defense**: Misbehavior is handled by protocol-level penalties: slashing or disqualification for double-signing or failure to deliver QKD keys. The hub-and-spoke QKD design allows redundant key paths to mitigate targeted outages.

### F. Quantum-Capable Signature Forgery

Quantum-capable signature forgery is a future threat in which big quantum computers could use Shor's algorithm to forge ECDSA or Ed25519 signatures that are stored on-chain. These kinds of enemies have quantum computing powers that let them carry out cryptographic attacks in polynomial time. QLink lessens this risk by only using NIST-standard post-quantum cryptographic signature schemes like Dilithium, Falcon, and SPHINCS+, which are made to stay safe from quantum threats.

The quantum capable signature forgery challenges represent as follows:

- **Threat:** Once large-scale quantum computers exist, adversaries can forge ECDSA/Ed25519 signatures recorded on-chain.

- *Adversary Capability:* Quantum adversary with polynomial-time access to Shor's algorithm.
- *QLink Defense:* All signatures in QLink are based on NIST-standard PQC schemes (Dilithium, Falcon, SPHINCS+), which are quantum-resistant.

## 3. Methodology

The design and evaluation of QLink follow a layered methodology that integrates cryptographic primitives, network simulation, and validator-level incentive modeling to demonstrate the feasibility of a quantum-safe interoperability protocol. This methodology reflects the multi-layered nature of QLink itself, combining post-quantum cryptography (PQC) for proof generation and key management, quantum key distribution (QKD) for secure validator communication, and hardware security modules (HSMs) for tamper-resistant key custody. Each layer was modeled, simulated, and analyzed to verify that quantum-resilient guarantees hold under realistic network and adversarial conditions.

The QLINK system uses a combination of classical and quantum-resistant technologies to safely link up different blockchains. The first step is the System Architecture, which uses Quantum Key Distribution (QKD) and Post-Quantum Cryptography (PQC) to keep communication safe, even from threats that may come up in the future. The Validator Network is the main part of the system. In this case, validators are trusted nodes that talk to each other over QKD links that are based on hubs and spokes or satellites. They are carefully enrolled, certified, and watched to make sure they follow the rules. If someone breaks the rules, strict rules are in place to keep the system safe. Finally, the Consensus Protocol and Proof Aggregation layer is where choices are made and checked. Validators work together to make sure that transactions are agreed upon, smart contracts are followed, and data across different chains, such as Bitcoin and Ethereum, is safe and consistent. All these layers work together to make QLINK a reliable bridge between different blockchain ecosystems as shown in fig 3. We have discussed Qlink methodology in detail at our below section.

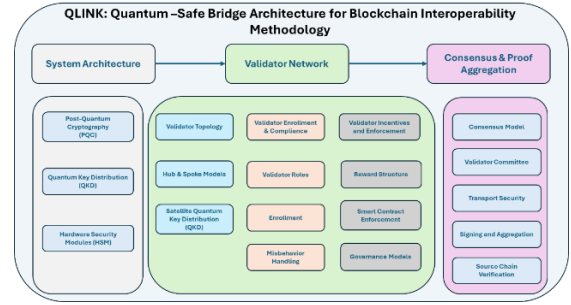

Fig. 3: Qlink: Quantum Safe Bridge Architecture for Blockchain Interoperability

### A. System Architecture

QLink is implemented as a dedicated **Layer-3 protocol** that operates above existing Layer-1 and Layer-2 blockchains. Its architecture integrates **two core cryptographic technologies** that together enable quantum-safe interoperability:

Quantum Key Distribution (QKD) establishes symmetric session keys between validators through quantum channels, providing information-theoretic security, tamper-evident key exchange, and perfect forward secrecy against both classical and quantum adversaries. Post-Quantum Cryptography (PQC) employs NIST-standardized digital-signature schemes such as Dilithium and Falcon, enforced within Hardware Security Modules (HSMs). Validator private keys remain sealed in hardware, ensuring that cross-chain signatures and proofs cannot be forged even under quantum attack.

In the PQC subsystem, each validator $v_i \in V$ maintains a key pair $(pk_i, sk_i)$ generated by a NIST-standardized PQC signature scheme such as Dilithium or Falcon. Private keys remain sealed inside Hardware Security Modules (HSMs), ensuring that signing operations never expose secret material. For a message m, the validator produces $\sigma_i = \text{Sign}(sk_i, m)$, and verification proceeds through $\text{Verify}(pk_i, m, \sigma_i)$. A quorum of validators forms an aggregated proof $\Pi = \{ \sigma_i : i \in I, |I| \geq T \}$, where I is the subset of validators that signed the message and T is the minimum threshold required for consensus, typically $T = 2f + 1$ for $n = 3f + 1$ validators.

The QKD subsystem generates symmetric keys between validators to encrypt and authenticate inter-validator messages. For each pair of connected validators $(v_i, v_j)$, the key generation rate $R_{ij}$ depends on the optical-fiber distance $d_{ij}$ according to the equation $R_{ij} = R_0 \cdot e^{-\lambda d_{ij}}$, where $R_0$ is the base rate and $\lambda$ represents channel attenuation. Let $B_{ij}$ be the validator traffic rate (bits per second). To preserve one-time-pad encryption and forward secrecy, QLink enforces $R_{ij} > B_{ij} \forall (v_i, v_j)$. Empirical simulations confirm this condition holds even at 50 km distances, with QKD producing 62–707× more key material than validator traffic. Keys are used once per message, providing information-theoretic confidentiality and rendering intercepted ciphertexts permanently indecipherable.

On top of these cryptographic foundations, Bridge Smart Contracts coordinate cross-chain operations. They lock and release assets between chains, verify PQC-signed proofs produced by QLink validators, and check that at least one active validator or regional hub operating QKD hardware is registered for the bridge. This requirement guarantees that every active bridge maintains a minimum quantum-secure communication path, preserving QLink's security baseline even if not all validators possess QKD equipment.

This modular architecture ensures that quantum-resilient guarantees are enforced from validator communication to contract verification, while remaining compatible with existing blockchain infrastructures (see Fig. 4).

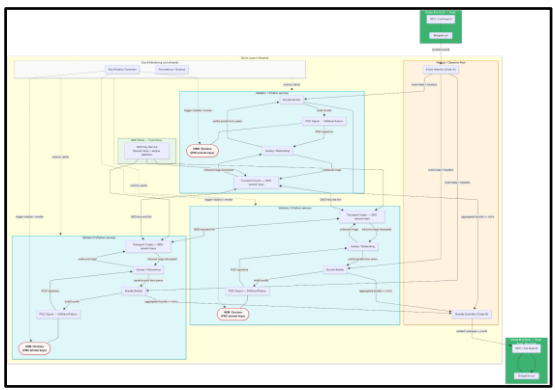

Fig. 4: QLink Architecture Diagram.

- ***The Mathematical Model for Qlink Architecture***

This modular architecture ensures that cryptographic guarantees are enforced across all layers, from user signatures to validator communication and contract verification. Building on the architectural and consensus mechanisms described above, this section formalizes QLink's internal composition through a mathematical abstraction. The model defines how the protocol's cryptographic, quantum, and consensus components interact to achieve quantum-safe cross-chain interoperability. By expressing these relationships analytically, we establish the theoretical foundation for the experimental simulations discussed in Section 3.

- ***Functional Representation:*** QLink can be represented as a composite function:

$$Q_{link} = f(PQC, QKD, V, C)$$

where each variable represents a subsystem within the Layer-3 protocol:

  o ***PQC*** – post-quantum cryptography module, including signature algorithms and hardware-enforced key custody.
  o ***QKD*** – quantum key distribution layer generating symmetric encryption keys between validators.
  o ***V*** – validator network defining node topology, weights, and thresholds.
  o ***C*** – consensus and cross-chain proof mechanism ensuring deterministic finality.

Together, these components form a closed system in which cryptographic and physical-layer security are mathematically linked.

- ***Security Condition and System Summary:*** QLink's operational stability depends on the dual condition:

$$R(d) > B \quad \text{and} \quad |I| \geq T = 2f + 1$$

The first inequality guarantees that quantum key generation exceeds communication demand, sustaining one-time-pad encryption. The second ensures that consensus finality requires super-majority approval, preserving safety and liveness even with up to one-third Byzantine validators.

In compact form, the overall system can be expressed as:

$$Q_{link} = PQC(HSM, Algo) + QKD(R(d)) + V(T) + C(BFT)$$

This model highlights how QLink integrates cryptographic, physical, and consensus-level assurances within a unified Layer-3 framework.

It establishes the theoretical basis for the experimental evaluation presented in Section 3, confirming that the protocol maintains operational efficiency, deterministic finality, and quantum-resilient security under realistic network conditions. The mathematical abstraction also provides a formal lens through which the next section examines QLink's security boundaries and adversarial resilience.

### B. Validator Network

Our simulations evaluate point-to-point QKD links between two validators. In practice, a production network must support committees of 21–31 validators distributed across regions. While the present design focuses on direct fiber links for metropolitan distances (≤50 km), future versions of QLink can extend to more complex topologies. Regional QKD hubs may relay keys to smaller validators, enabling broader participation without requiring each validator to maintain direct fiber links. For intercontinental communication, satellite-based QKD has already been demonstrated at >1000 km, and QLink can integrate such channels as the technology matures. This forward-looking design ensures QLink remains scalable beyond point-to-point setups while remaining feasible with current technology.

QLink does not rely on token-based governance or Decentralized Autonomous Organization (DAO) style voting. Instead, validator participation is managed at the protocol level through registration in Layer-3 bridge contracts. Not all validators require direct QKD hardware. A subset of certified validators, or regional QKD hubs, are responsible for generating and distributing quantum keys, while other validators can securely consume these keys for consensus traffic. Only the validators (or hubs) that operate QKD transmitters/receivers must provide proof of certification and compliance, ensuring that the key generation plane maintains its quantum-safe guarantees. Validators (with or without hardware) register their PQC public keys in the bridge contracts, and consensus recognizes only signatures from registered keys. Validators that double-sign, submit invalid proofs, or fail to maintain required connectivity can be slashed or removed from the whitelist. Misconduct at the QKD layer (e.g., a certified hub failing to deliver keys) leads to disqualification or reallocation of responsibilities. This role-based design allows QLink to remain practical: a few certified QKD-equipped validators or hubs can support a larger validator set, while the protocol still ensures that every signature and message is secured by post-quantum cryptography and backed by QKD-generated session keys.

Operating certified Quantum Key Distribution (QKD) devices involves higher upfront costs compared to classical bridge validators. To balance these costs and encourage participation, QLink introduces a reward and enforcement framework that differentiates between QKD-equipped hubs and general validators. Validators earn transaction fees (percentage of each transfer), optional block rewards in the form of minted bridge tokens, staking yield from bonded tokens, and QKD subsidies to offset hardware investment. In addition, reputation boosts make validators in a quantum-secure bridge more attractive for institutional partnerships. Validator eligibility is controlled by a ValidatorRegistry contract, with only validators possessing verified QKD hardware certificates being whitelisted. Certificates are issued by trusted hardware vendors and linked to validator addresses via on-chain proofs. Bridge contracts accept signatures exclusively from registered validators, preventing unauthorized participation. QLink supports multiple validator selection models: permissioned validators curated by governance, delegated proof-of-stake with QKD compliance checks, and hybrid models where validator hubs operate QKD links on behalf of smaller relayers. These models ensure flexibility while preventing centralization.

The validator network is defined as $V = \{v_1, v_2, \ldots, v_n\}$. Each validator carries a weight $w_i$ (e.g., stake or equal vote), and total weight $W = \Sigma w_i$. A consensus round succeeds when the cumulative weight of participating validators satisfies $\Sigma w_i \geq T$ where $T > \frac{2}{3} W$, ensuring Byzantine Fault Tolerance. The subset $I \subseteq V$ represents validators that validated and signed a given message m. When $|I| \geq T$, m is finalized and yields a cross-chain proof $(m, \Pi_m)$.

### C. Consensus Protocol and Proof Aggregation

To ensure deterministic safety in cross-chain messaging, QLink employs a Byzantine Fault Tolerant (BFT) consensus model adapted for a quantum-secure validator network. The protocol achieves rapid finality (typically within 1–3 seconds per round) while leveraging stake-based Sybil resistance, quantum-secure communication, and post-quantum cryptography.

Validators participate in a stake-weighted BFT protocol (inspired by Tendermint / HotStuff), where a rotating leader proposes bundles of cross-chain events. Finality is achieved once 2f+1 of n validators sign, ensuring safety and liveness under adversarial conditions. Typical committee size ranges from 21–31 globally distributed validators, tolerating up to $f = \lfloor(n-1)/3\rfloor$ Byzantine faults. Leader rotation occurs at fixed epochs or slots using verifiable randomness. All consensus messages are encrypted with session keys derived from the QKD plane, guaranteeing confidentiality and forward secrecy against both classical and quantum adversaries.

Validators sign votes and bundles using PQC algorithms (ML-DSA/Dilithium by default, Falcon or SPHINCS+ as alternatives). Aggregation may employ threshold-PQC signatures where available or fallback to t-of-n multisignature schemes, with all private keys confined to HSM enclaves. QLink validators independently verify source-chain events prior to consensus. For Bitcoin, validators maintain a Simplified Payment Verification (SPV) header chain and verify Merkle proofs of lock events, with k-confirmations as a configurable security parameter. For Ethereum, validators track finalized blocks through Remote Procedure Call (RPC) quorums or verify light-client proofs (sync committees), ensuring events are validated before cross-chain finality.

By combining deterministic finality, quantum-secure transport, and post-quantum signatures, and by leveraging the QKD surplus demonstrated in Section 3, QLink delivers cross-chain proofs that are resistant to both classical exploits and emerging quantum threats, while keeping validator overhead negligible relative to blockchain confirmation times.

*4.3.1 Consensus and Proof Verification*

On the destination chain, the bridge contract verifies the aggregated proof only if every signature $\sigma_i$ in $\Pi_m$ is valid under its corresponding $pk_i$, all signers are registered ($pk_i \in V_{(reg)}$), and the quorum threshold $|I| \geq T$ is satisfied. If all conditions hold, the contract executes the associated mint or release operation. This ensures that no minority subset can forge or finalize a cross-chain message

## 4. Experiment and Results

To evaluate the feasibility of QLink's quantum-secure validator network, we conducted simulations of QKD-based key exchange across varying fiber-optic link distances: 5 km, 10 km, 50 km. The objective was to determine whether QKD-generated session keys could support validator communication without introducing significant overhead to cross-chain transactions. As a motivating application, we focused on the Bitcoin → Ethereum bridge, the most widely used and most frequently exploited interoperability path in Web3. In addition to the base two node experiments, we also ran extended simulations with validator committees of 7, 8, and 9 nodes. These tests used the same link distances, traffic profile, and cryptographic setup.

### 4.1. Experimental Setup

The OpenQKD simulator, built on an extended version of QuNetSim/QKDNetSim, models a secure communication setup between two validators linked through a fiber-optic QKD channel with a 50 Mbit buffer. It employs VMAC authentication and one-time pad (OTP) encryption to ensure that every validator message uses a fresh cryptographic key. The network handles a steady 20 kbps of application traffic, simulating consensus and proof-signing exchanges for a Bitcoin-to-Ethereum bridge, benchmarked against Bitcoin's 60-minute and Ethereum's 13-minute confirmation times.

- **Simulator**: OpenQKD simulator (extended from QuNetSim/QKDNetSim).
- **Configuration**: Two validators connected by a fiber-optic QKD channel with a 50 Mbit buffer capacity.
- **Security Layer**: VMAC authentication and one-time pad (OTP) encryption, ensuring fresh key usage for every validator message.
- **Application Traffic**: Fixed at 20 kbps, representing consensus and proof-signing traffic for a Bitcoin → Ethereum bridge.
- **Baseline**: Bitcoin confirmation (~60 minutes) and Ethereum confirmation (~13 minutes).

### 4.2. Results

Across all distances, the system performed efficiently with minimal packet loss (<0.1%). Over the 5 km link, it achieved about 13.1 Mbit/s, generating roughly 656 million bits in 50 seconds, over 700 times more than needed for validator traffic. At 10 km, the rate averaged 10.3 Mbit/s, yielding about 515 million bits and maintaining a 550× surplus. Even at 50 km, where the rate dropped to 1.16 Mbit/s, it still produced around 58 million bits, more than 60 times the required amount. Results has describe in below based on distances:

- **5 km link**: Achieved average key rate of ~13.1 Mbit/s, generating ~656 million bits in 50 seconds. Validator traffic consumed ~930 Kb, leaving a ~707× surplus of keys. Only 2 missed packets recorded (<0.1% loss).
- **10 km link**: Achieved average key rate of ~10.3 Mbit/s, generating ~515 million bits in 50 seconds. Consumption remained ~930 Kb, leaving a ~550× surplus of keys. Only 1 missed packet recorded.
- **50 km link**: Key rate dropped to ~1.16 Mbit/s, generating ~57.8 million bits in 50 seconds. Still exceeded validator consumption by ~62×, with just 1 missed packet (<0.1% loss).

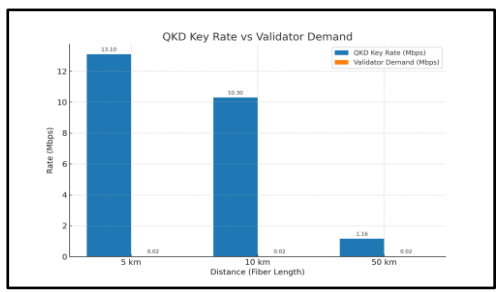

Fig. 5. QKD Key Rate vs Validator Demand or Bitcoin → Ethereum bridge traffic

### 4.3. Interpretation

The results show that distance scaling had a predictable impact on performance, with key rates decreasing as distance increased. However, even at 50 km, the system still generated far more keys than the validator required, showing excellent efficiency. Validator consumption remained minimal across all tests, confirming low operational overhead. Reliability was strong, with almost no packet loss, and the large surplus of keys provides strong security guarantees for sustained, secure communication. Here are the major intrpreation of Qlink as shown in fig. 5.

- **Distance Scaling**: Shorter fiber distances produced higher key rates: ~13.1 Mbps at 5 km versus ~1.16 Mbps at 50 km. While attenuation reduces throughput, QKD generated key material far above application demand at every distance.
- **Validator Consumption:** Application traffic consumed ~930 Kb of key material, while generated keys were between 62× and 707× greater than required. This confirms validators will not exhaust QKD material even under high network load in a Bitcoin → Ethereum bridge.
- **Reliability:** Key/data utilization was ~99–100%, with only 1–2 missed packets (<0.1% impact). Validator-to-validator communication remained stable throughout all runs.
- **Security Guarantees**: OTP encryption provided information-theoretic confidentiality, while VMAC authentication prevented message forgery or tampering. Combined with HSM-enforced PQC signatures, this ensures validator messages are quantum-safe end to end.

### 4.4. Implications for Qlink

For QLink, the results suggest minimal cryptographic overhead since key generation far exceeds usage, reducing bottlenecks in encryption processes. The small transaction size further limits processing demand, improving overall efficiency. With fast key availability and low validator traffic, end-to-end latency is expected to stay low, supporting real-time, secure communication even over longer distances. PQC signing inside HSMs (Dilithium, Falcon) added <10 ms per validator, negligible compared to blockchain confirmation times. Proof bundles increased from ~1 KB (ECDSA) to ~3–6 KB under PQC, a modest cost. Bitcoin (~60 min) and Ethereum (~13 min) dominate total bridge latency (~73 min), while QKD, PQC, and HSM enforcement add <1 second of overhead. Thus, QLink achieves quantum-safe bridging without reducing throughput or introducing meaningful latency. Future work extends simulations to 1000 km using satellite-based QKD, validating global-scale deployment.

### 5. Conclusion and Future Work

This paper introduced QLink, a Layer-3 interoperability protocol that secures cross-chain bridges with quantum key distribution (QKD), post-quantum cryptography (PQC), and hardware-enforced key custody (HSMs). We identified validator compromise as the central weakness in existing bridge designs and proposed a validator architecture where all communication is quantum-secure and cryptographically robust against future quantum adversaries. Through simulations at 5 km, 10 km, and 50 km fiber distances, we demonstrated that QKD throughput consistently exceeded validator consumption by 62× to 707×, even under long-haul conditions. This confirms that QKD does not introduce latency or throughput bottlenecks. Combined with PQC signatures and HSM enforcement, QLink adds <1 second of cryptographic overhead to cross-chain proofs, leaving total transaction time dominated by blockchain confirmation (e.g., ~60 minutes for Bitcoin, ~13 minutes for Ethereum). QLink therefore enables quantum-safe Bitcoin ↔ Ethereum transfers today, while laying the foundation for secure interoperability across the multi-chain ecosystem.

Future work will extend our simulations to larger validator committees, explore hub-and-spoke QKD topologies and satellite QKD for global coverage, and evaluate performance under adversarial network conditions. By combining cryptographic rigor with practical scalability, QLink offers a path to future-proof cross-chain bridges against both present-day exploits and emerging quantum threats.